\documentclass[apl,twocolumn,preprintnumbers,amsmath,amssymb,floatfix]{revtex4-1}
\usepackage{graphicx}
\usepackage{lmodern}
\DeclareUnicodeCharacter{0308}{HERE!HERE!}
\DeclareUnicodeCharacter{0301}{***}
\usepackage{tabularx}
\usepackage{siunitx}
\usepackage{booktabs}
\usepackage{etoolbox}
\usepackage{epstopdf}
\usepackage{float}
\usepackage{placeins}
\usepackage{hyperref,graphicx}
\usepackage{dcolumn}
\usepackage{bm}
\usepackage{color}
\usepackage{natbib}
\usepackage{amsmath}
\usepackage{amssymb}
\usepackage{multirow} 
\usepackage{hyperref}
\usepackage{stackengine}
\usepackage{xcolor}
\hypersetup{colorlinks=true, linkcolor=blue, citecolor=red, urlcolor=magenta, pdftitle={DDPOs}, pdfauthor={Saurabh Ghosh}}
\begin{document}
\title{Low-Energy Magnetic States of Tb Adatom on Graphene 
}
\author{Monirul Shaikh}
\author{Alison Klein}
\author{Aleksander L. Wysocki}
\email{wysockia@unk.edu}
\affiliation{Department of Physics and Astronomy, University of Nebraska at Kearney, Kearney, NE 68849, United States}

\begin{abstract}
Electronic structure and magnetic interactions of a Tb adatom on graphene are investigated from first principles using combination of density functional theory and multiconfigurational quantum chemistry techniques including spin-orbit coupling. We determine that the six-fold symmetry hollow site is the preferred adsorption site and we investigate electronic spectrum for different adatom oxidation states including Tb$^{3+}$, Tb$^{2+}$, Tb$^{1+}$, and Tb$^{0}$. For all charge states, the Tb $4f^8$ configuration is retained with other adatom valence electrons being distributed over $5d_{xy}$, $5d_{x2+y2}$, and $6s/5d_0$ single-electron orbitals. We find strong intra-site adatom exchange coupling that ensures that the $5d6s$ spins are parallel to the $4f$ spin. For Tb$^{3+}$, the energy levels can be described by the $J=6$ multiplet split by the graphene crystal field. For other oxidation states, the interaction of $4f$ electrons with spin and orbital degrees of freedom of $6s5d$ electrons in the presence of spin-orbit coupling results in the low-energy spectrum composed closely lying effective multiplets that are split by the graphene crystal field. Stable magnetic moment is predicted for Tb$^{3+}$ and Tb$^{2+}$ adatoms due to uniaxial magnetic anisotropy and effective anisotropy barrier around 440 cm$^{-1}$ controlled by the temperature assisted quantum tunneling of magnetization through the third excited doublet. On the other hand, in-plane magnetic anisotropy is found for Tb$^{1+}$ and Tb$^{0}$ adatoms. Our results indicate that the occupation of the $6s5d$ orbitals can dramatically affect the magnetic anisotropy and magnetic moment stability of rare earth adatoms. 
\end{abstract}
\maketitle
\section{Introduction}

Magnetic adatoms has recently garnered significant interest due to their potential applications in spintronics and quantum information technologies\cite{Donati2021}. Individual magnetic adatoms on surfaces may exhibit long-lived spin quantum states and function as single atom magnets with significant remanent magnetization and coercivity at temperatures as high as several tens of kelvin\cite{Donati2016}. Reading and writing\cite{Natterer2017} as well as coherent control\cite{Yang2018} of the adatoms spin states have been demonstrated making such systems promising materials for ultrahigh-density magnetic information storage and quantum logic devices.

Rare-earth adatoms carrying a nonzero orbital angular momentum are particularly interesting since their strong spin-orbit coupling (SOC) may lead to a large magnetic anisotropy barrier which is essential to obtain robust adatom magnetic moment that is stable against fluctuations from the environment. Different late rare-earth-adatoms have been realized using various substrates including metallic (Cu, Pt) surfaces\cite{Donati2014,Singha2017}, insulating surfaces like MgO\cite{Donati2016,Natterer2017,Natterer2018,Singha2021,Singha2021b} or BaO\cite{Sorokin2023}, and graphene\cite{Baltic2016,Baltic2018}. While lanthanide adatoms on metals may have a large magnetic anisotropy barrier, effective scattering with conduction electrons and soft phonon modes of the metallic surface lead to efficient under-the-barrier relaxation processes which make the adatom magnetic states short-lived\cite{Donati2014}. In the case of insulating surfaces, like MgO, however, the lack of conduction electrons and low phonon density of states result in long magnetic relaxation time and magnetic hysteresis provided that the adatom site symmetry leads to highly axial magnetic states and suppresses quantum tunneling of magnetization\cite{Donati2016,Natterer2018}. Similarly, stable magnetic moments were observed for Dy adatoms on graphene thanks to low substrate electron and phonon densities and the high symmetry of the adatom adsoprtion site\cite{Baltic2016,Baltic2018}. Graphene is, in fact, especially promising substrate due to absence of nuclear spins in abundant (98.9\%) $^{12}$C nuclei. Since hyperfine coupling with nuclear spins is one of the main decoherence mechanisms, one may expect adatoms on graphene to have a longer coherence times which is of vital importance for potential quantum information applications.

Magnetic adatoms share a lot of physics with single molecular magnets (SMMs)\cite{SMMbook}. Even though the former have no ligands, the crystal field (CF) is instead provided by the substrate. For both systems, the interplay of the CF and SOC is responsible for magnetic anisotropy. The lack of ligands for adatom systems can actually be considered advantageous since the interactions with ligand vibrational modes is one of the main magnetic relaxation mechanisms in SMMs\cite{SMMbook}. An important point is that lanthnide-based SMMs are typically in the trivalent  oxidation state while rare-earth adatoms may adapt various oxidation states. In fact, the adatom oxidation can be potentially controlled externally, for example, by gating as demonstrated for Co adatoms on graphene\cite{Brar2011}. Different lanthanide oxidations may have different electronic excitation spectrum and, therefore, different magnetic properties like magnetic anisotropy or hyperfine coupling\cite{Smith2020}. Consideration of various oxidation states is, thus, crucial in order to elucidate magnetic properties of rare-earth adatoms.

Many theoretical studies of lanathanides adatoms are based on effective spin Hamiltonian\cite{Natterer2018,Baltic2018,Curcella2023} or ligand field models\cite{Donati2016,Baltic2018} to describe low-lying magnetic states and anisotropy barrier. While such approaches are often very useful to explain experimental data, they involves many free parameters and cannot always provide a clear physical description of the system. On the other hand, first principles calculations, while parameter-free, are typically based on density functional theory (DFT)\cite{Donati2016,Pivetta2020,Donati2020,Singha2021} which cannot properly describe the multireference nature of the lanthanide $4f$ electrons. \emph{Ab initio} quantum chemistry techniques has proven to be very successful in describing lanthanide SMMs\cite{CompNanoMAgBook} and, therefore, is an ideal method to study rare-earth adatoms\cite{Dubrovin2021}.

Here, we present our first principles studies of electronic structure and magnetic properties of Tb adatom for different oxidation states. Preferred adsorption sites and atomic coordinates are determined using DFT calculations in the supercell geometry. Next, electronic spectrum and magnetic interactions are studied using multiconfigurational quantum chemistry methods for a cluster model. The intra-site exchange coupling, effective magnetic moments, CF parameters, and effective anisotropy barrier for different charge states of the Tb adatoms are discussed.

\section{Methodology}

Adsorption energy calculations and atomic relaxations for a Tb adatom on a graphene for different adsorption sites are performed using density functional theory (DFT) with PBE exchange-correlation functional\cite{PBE}. The DFT calculations are done using a supercell method with the Tb adatom on the $8\times8$ graphene supercell (128 C atoms). The graphene layer is separated from its periodic image by 15 ${\AA}$. The Kohn-Sham equations are solved using the projector augmented wave (PAW) method\cite{kresse1999ultrasoft} as implemented in the VASP code\cite{vasp}. Since DFT cannot properly describe partially filled $4f$ states, for Tb we use the open-core PAW pseudopotential (corresponding to the $4f^8$ configuration). As this approach does not include the $4f$ exchange-correlation potential when computing valence electrons, the calculations are non-spin polarized. The cutoff energies for the plane wave and augmentation charge are 500 eV and 645 eV, respectively. We use $\Gamma$-centered 4$\times$4$\times$1 $k$-point mesh. Atomic positions are relaxed until the Hellmann-Feynman forces are converged to less than 0.01 eV/${\AA}$.

\begin{figure}
\centering
\includegraphics[width=\linewidth]{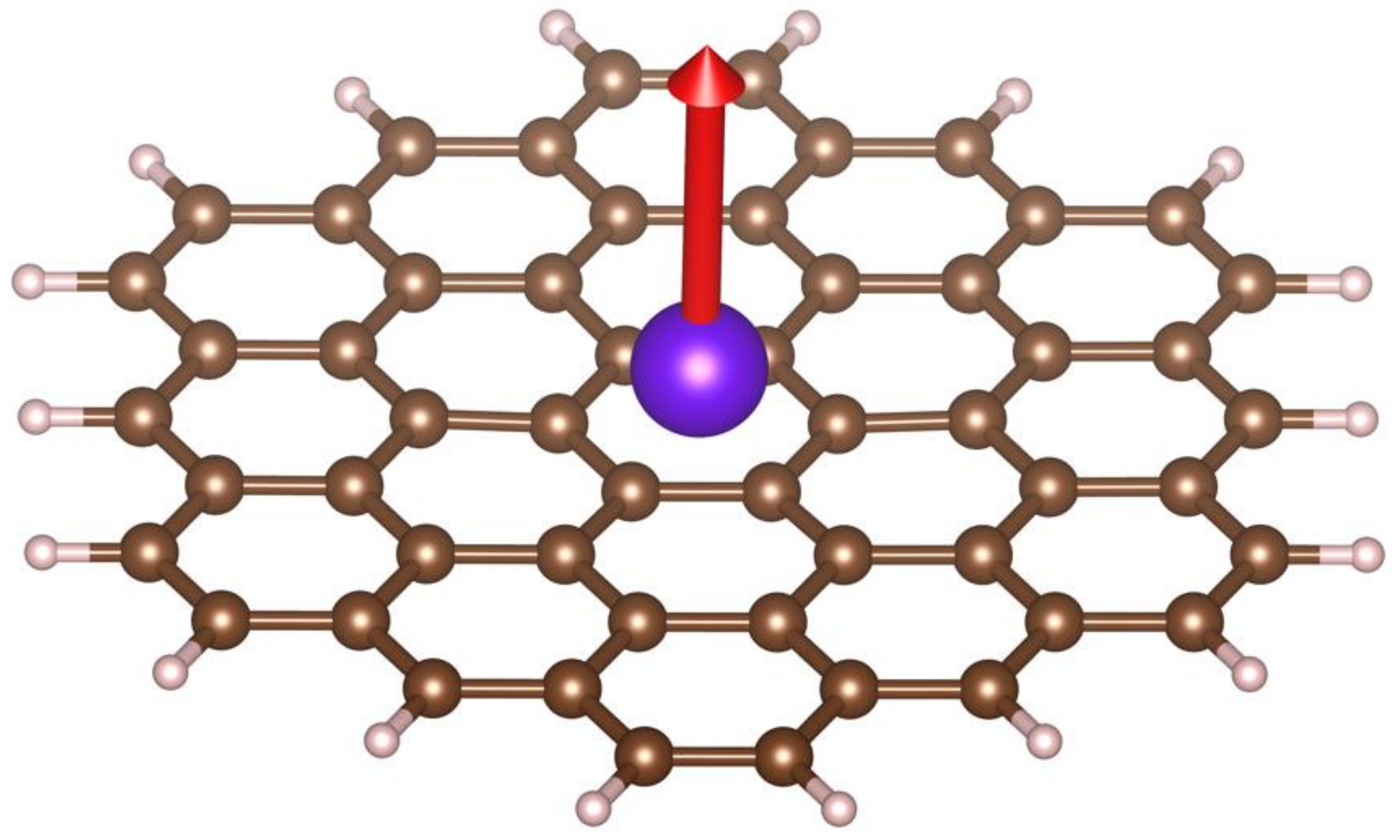}
\caption {Cluster used for quantum chemistry calculations of Tb adatom on graphene. Purple, brown, and light red spheres represent terbium, carbon, and hydrogen atoms, respectively. The red arrow represents the adatom magnetic moment. The Tb adatom is exactly on top of the graphene hollow site so that the cluster retains the $C_{6v}$ symmetry.} 
\label{Cluster}
\end{figure}

For quantum chemistry calculations we use a cluster model constructed from the discussed above DFT-optimized supercell. The cluster consists of the Tb adatom and 54 neighboring C atoms. The carbon dangling bonds are passivated by hydrogen atoms using the 1.09 $\AA$  C-H bod lengths such that the $C_{6v}$ symmetry is retained. The resulting cluster is shown in Fig.~\ref{Cluster}. A somewhat smaller cluster was used for quantum chemistry calculations of Co adatom on graphene\cite{Rudenk2012}. Note that the approximation of the infinite graphene layer by a finite cluster dos not allow to fully reproduce the graphene band structure including the Dirac cone. However, our focus here is on the proper description of the local excitation spectrum of the Tb adatom. These energy levels are primarily affected by the adatom-graphene interaction which should be well described with the cluster model. Our approach is similar to quantum chemistry calculations of the negatively charged nitrogen-vacancy (NV$^-$) center in diamond\cite{Bhandari2021}. In that work a finite cluster was adapted to represent a diamond lattice and even though the diamond band structure was not perfectly described, the local excitation spectrum of the NV$^-$ center was found to be in a good agreement with experiment\cite{Bhandari2021}.

The multiconfigurational quantum chemistry calculations are performed using the OpenMolcas code\cite{Openmolcas}. The adatom oxidation state is selected by adjusting the total charge of the cluster. Scalar relativistic effects are included based on the second order Douglas-Kroll-Hess Hamiltonian\cite{douglas1974quantum,Hess1986} and relativistically contracted atomic natural orbital (ANO-RCC) basis sets\cite{widmark1990,roos2004}. We use polarized valence triple-$\zeta$ (ANO-RCC-VTZP) basis set for the Tb atom, polarized valence double-$\zeta$ (ANO-RCC-VDZP) basis set for C atoms, and valence double-$\zeta$ (ANO-RCC-VDZ) basis set for H atoms. The calculations are performed in two steps. First,in the absence of spin-orbit coupling (SOC), the spin-free energies and eigenstates are evaluated for a given spin multiplicity using the state-averaged complete active space self-consistent field (SA-CASSCF)\cite{roos1980b, siegbahn1981}. The active space, spin multiplicities, and number of roots used in SA-CASSCF calculations depend on the Tb adatom oxidation state and will be discussed in section 4.
In the second step, the SOC is included within the atomic mean-field approximation\cite{hess1996mean} using the restricted active space state interaction (RASSI) method\cite{malmqvist2002}. The RASSI-SOC calculations are performed within the space spanned by all considered spin multiplicities and SA-CASSCF roots (including their spin-sublevels). CF parameters and total angular momenta components for electronic doublets are calculated using the  SINGLE$\_$ANISO formalism\cite{Chibotaru2012}.

\section{DFT Calculations}

\begin{table}
\begin{center}
\caption{\label{tab:1} DFT calculated adsorption energy and the optimized vertical distances of the Tb adatom from the graphene layer for different adsorption sites. The adsorption energies are relattive to the hollow site.}
\begin{tabular}{c|c|c}
 \hline
Adsorption & Adsorption  & Vertical \\
site & energy (meV) & distance (\AA)\\
   \hline
 Hollow site & 0 & 2.28 \\
 Top site & 85 & 2.32 \\
 Bridge site & 91 & 2.38 \\
 \hline
\end{tabular}
\end{center}
\end{table}

\begin{figure}
\centering
\includegraphics[width=\linewidth]{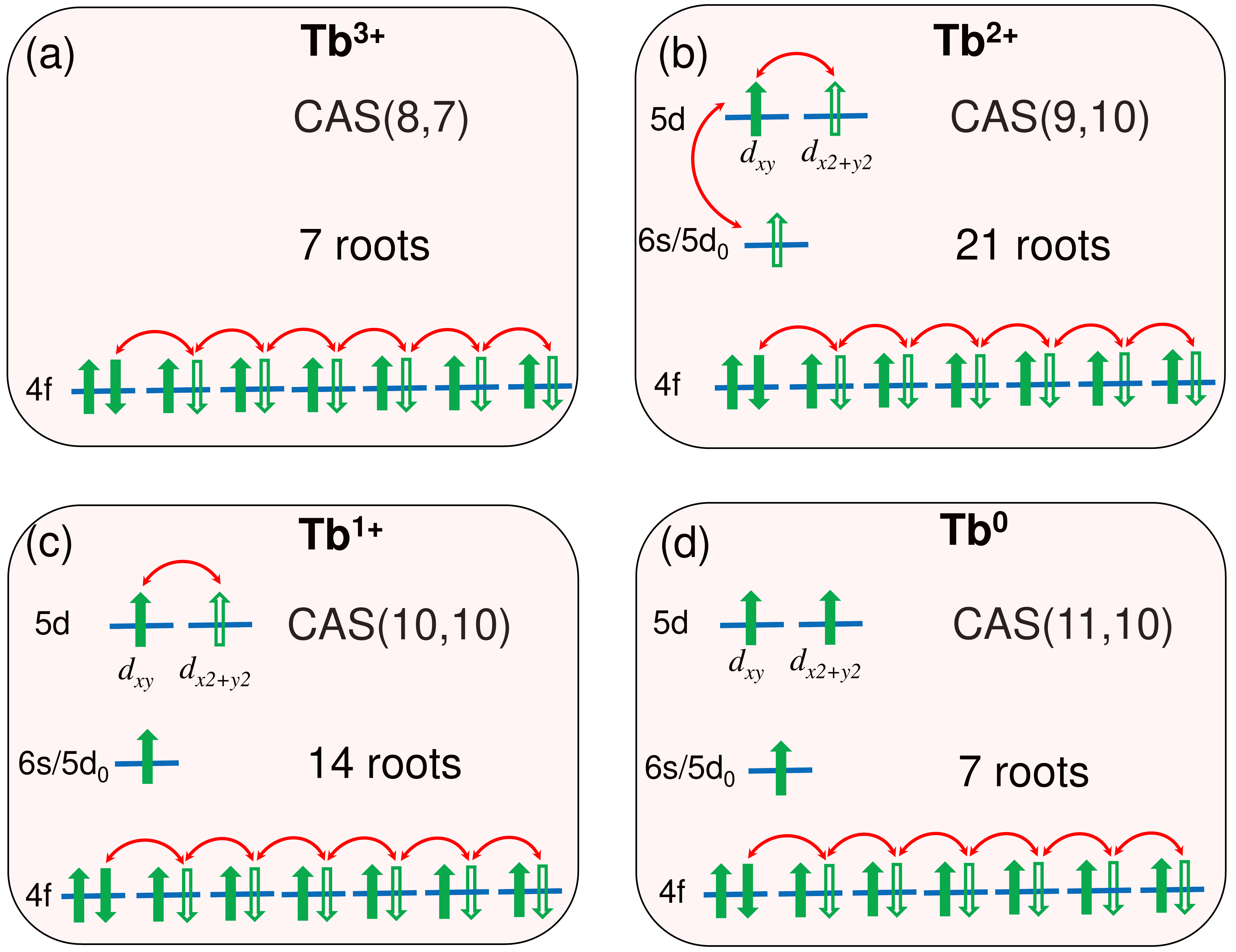}
\caption {Schematic diagram illustrating the selected active space and relevant configurations that determine number of roots used in SA-CASSCF calculations for different adatom oxidation states. We show only the case of the highest considered spin multiplicity. For a given oxidation states, the active space and number of roots are the same for all considered spin multiplicities. (a) For Tb$^{3+}$, the active space consists of eight electrons in seven $4f$-like orbitals with $S_{4f}=3$. We include seven roots that originate from seven possible configurations of the spin down electron in $4f$ orbitals. (b) For Tb$^{2+}$, for low-lying roots, the additional electron occupies a $6s/5d_0$ orbital hybrid and the pair of $d_{xy}$, $d_{x2+y2}$ orbitals. These three orbitals and the extra electron are, thus, added to the active space. As the extra electron has three possible configurations, the number of roots is $3\times7=21$. (c) For Tb$^{1+}$, an additional electron is added to the active space, but the number and type of active orbitals remain the same. For low lying roots, one of the non-$4f$ electrons resides at the $6s/5d_0$ hybrid while the other occupies $d_{xy}$- and $d_{x2+y2}$-like orbitals so that it has two possible configurations and the number of roots is $2\times7=14$. (d) For Tb$^{0}$, an additional electron is added to the active space, but the number and type of active orbitals again remain the same. Since we have three electrons occupying $6s/5d_0$, $d_{xy}$, and $d_{x2+y2}$ orbitals, this results in a single configuration and the number of roots is 7.} 
\label{LevelDiagram}
\end{figure}

In order to determine the preferred adatom-substrate geometry we consider three possible adsorption sites: i) hollow site with the adatom on top of the center of a single graphene honeycomb ($C_{6v}$ symmetry); ii) top site with the adatom on top of a carbon atom ($C_{3v}$ symmetry); iii) bridge site with the adatom on top of the center of a nearest-neighbor graphene bond ($C_{2v}$ symmetry). For each adsorption site we optimize atomic positions and calculate the total energy. Table 1 shows calculated adsorption energies and the vertical distance of the Tb adatom from the graphene layer. In general, we find that the Tb adatom is about 2.3-2.4~\AA~above the graphene layer and that this distance is the smallest for the hollow site. The hollow site has also the lowest energy with the energies of other sites being almost 0.1 eV larger. This indicates that the six-fold symmetric hollow site is the preferred adsorption site. This is consistent with scanning tunneling microscopy measurements of Dy adatoms on graphene that found that the adatoms are adsorbed at the hollow site\cite{Baltic2016}. 


\section{Quantum Chemistry Calculations}

Quantum chemistry calculations are performed for a cluster with the Tb adatom at the hollow site as shown in Fig.~\ref{Cluster}. Electronic structure and magnetic interactions are computed for different Tb oxidation states including Tb$^{3+}$, Tb$^{2+}$, Tb$^{1+}$, and Tb$^{0}$.

\subsection{Tb$^{3+}$ oxidation}

Tb adatom in the $+3$ oxidation state has the $4f^8$ valence electronic configurations with the Tb $6s$, $5d$, and $6p$ orbitals being unoccupied. In order to describe the strong electronic correlations of the partially filled $4f$ shell the SA-CASSCF active space consists of eight electrons and seven $4f$-like orbitals. Such CAS(8,7) active space was used to describe magnetic properties of the anionic TbPc$_2$ SMM where the Tb atom is also in the $+3$ oxidation state\cite{Wysocki2020}. According to the Hund's fist rule, the lowest energy states correspond to the maximum spin of the $4f$ electrons, i.e., $S_{4f}=3$. Other spin states are significantly higher in energy. Since the graphene layer is nonmagnetic and the adatom has no non-$4f$ electrons, the SA-CASSCF calculations are done for the total spin $S_{tot}=S_{4f}=3$. The spin state and the active space used for the Tb$^{3+}$ adatom is shown schematically in Fig.~\ref{LevelDiagram}a. The active orbital pictures are shown in Fig.~S1 in Supplementary Materials. Note that the spin down electron may occupy any of the seven $4f$ orbitals. This gives rise to seven possible spin-free states (roots) for the $4f^8$ and $S_{4f}=3$ configuration. Indeed, this configuration leads to the $4f$ orbital angular momentum of $L_{4f}=3$ and the seven roots correspond to the seven possible values of the magnetic orbital quantum number $M_L$. For a spherically symmetric isolated Tb$^{3+}$ ion, these roots would be degenerate. For the adatom geometry, however, this degeneracy is removed by the graphene CF. Nevertheless, as the CF acting on $4f$ electrons is relatively weak, the roots lie very close in energy and can be mixed by SOC. Therefore, all seven roots are included in the SA-CASSCF calculations and the corresponding state averaging procedure.

\begin{figure}
\centering
\includegraphics[width=\linewidth]{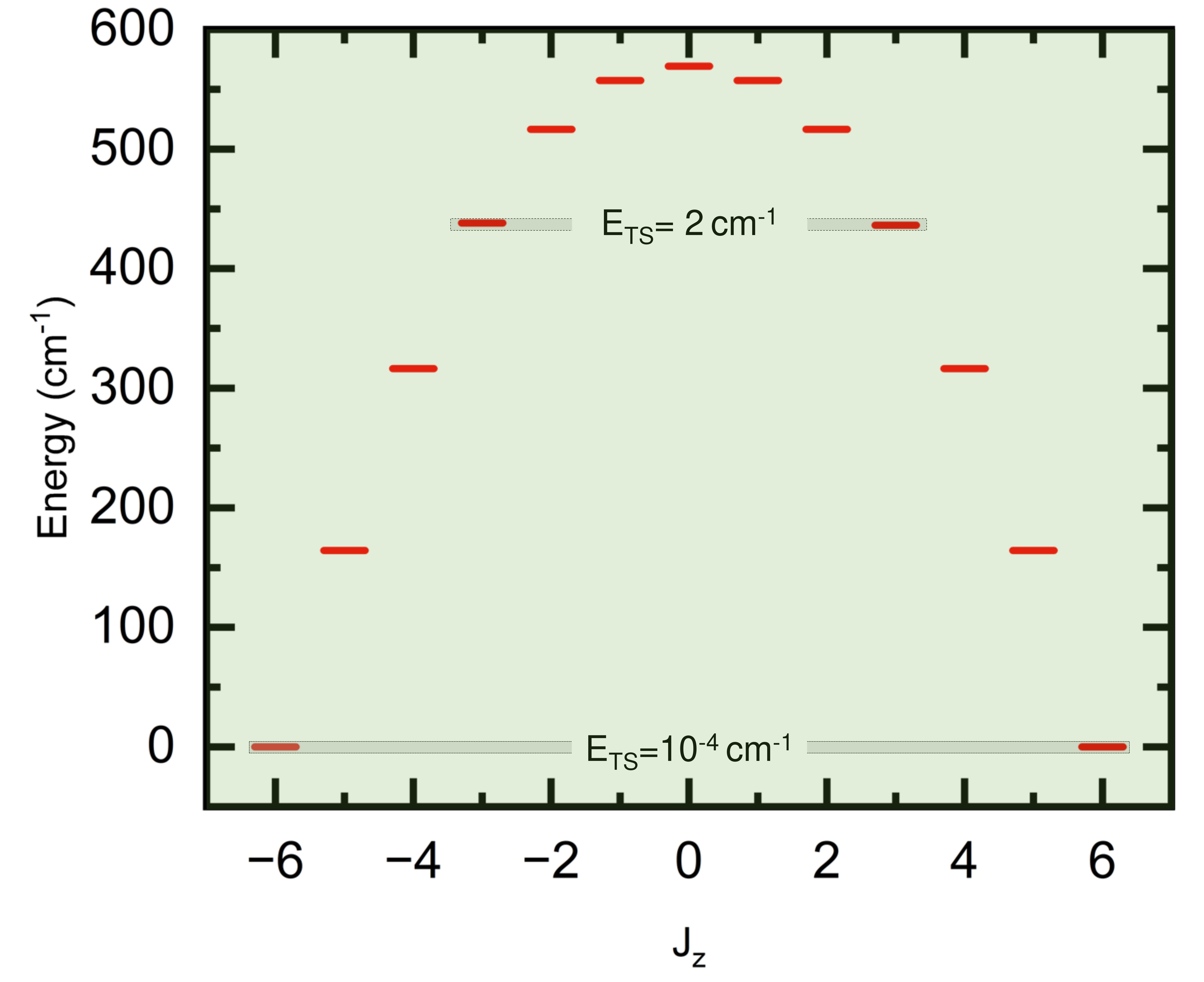}
\caption {Calculated low-energy magnetic states for the Tb$^{3+}$ adatom on graphene. The spectrum is primarily  composed of (quasi-)doublets. Each doublet state is placed along the horizontal axis according to its value of the $z$ component of the total angular momentum. The only singlet state has $J_z=0$. Tunnel splittings of the doublets due to $B_6^6$ transverse CF interaction are indicated. 
} 
\label{Q3+Energy}
\end{figure}

SOC mixes the $S_{4f}=3$, $L_{4f}=3$ states and gives rise to a multiplet structure with each multiplet corresponding to a different possible value of the $4f$ total angular momentum, $J$. It follows from the Hund's third rule that the lowest energy multiplet has the maximum possible total angular momentum of $J=6$. This ground multiplet is split by the graphene CF producing the calculated low-energy spectrum shown in Fig.~\ref{Q3+Energy} (the numerical values are shown in Table S1 in Supplementary Materials). The energy difference between the lowest and highest electronic levels within the ground multiplet is $E_{MAB}=569$ cm$^{-1}$. Despite being split, the ground multiplet remain well-defined as it is separated from higher lying levels by 1416 cm$^{-1}$ (not shown in Fig.~\ref{Q3+Energy}, see Table S1 in Supplementary Materials).

\begin{table}[b]
\begin{center}
\caption{\label{tab:2} Principal values of the $g$-tensor ($g_i$) and the CF parameters ($B_q^k$) calculated for the low-lying magnetic states of the Tb adatom on graphene for different adatom oxidation states. The units of CF parameters are cm$^{-1}$.}
\begin{tabular}{c|cccc}
 \hline
& Tb$^{3+}$  & Tb$^{2+}$ & Tb$^{1+}$  & Tb$^{0}$\\
   \hline
$g_{1}$ & 1.508     & 1.503     & 2.511     & 1.872 \\
$g_{2}$ & 1.508     & 1.503     & 2.511     & 1.872 \\
$g_{3}$ & 1.480     & 1.435     & 2.022     & 1.668 \\
$B_2^0$ & -5.375455 & -3.940770 &  6.071690 &  7.641457 \\
$B_4^0$ &  0.000216 & -0.005776 &  0.022960 &  0.004203 \\
$B_6^0$ &  0.000021 &  0.000001 & -0.000007 &  0.000005 \\
$B_6^6$ &  0.000041 &  0.000027 &  0.000139 &  0.000003 \\
\hline
\end{tabular}
\end{center}
\end{table}

\begin{figure}[b]
\centering
\includegraphics[width=\linewidth]{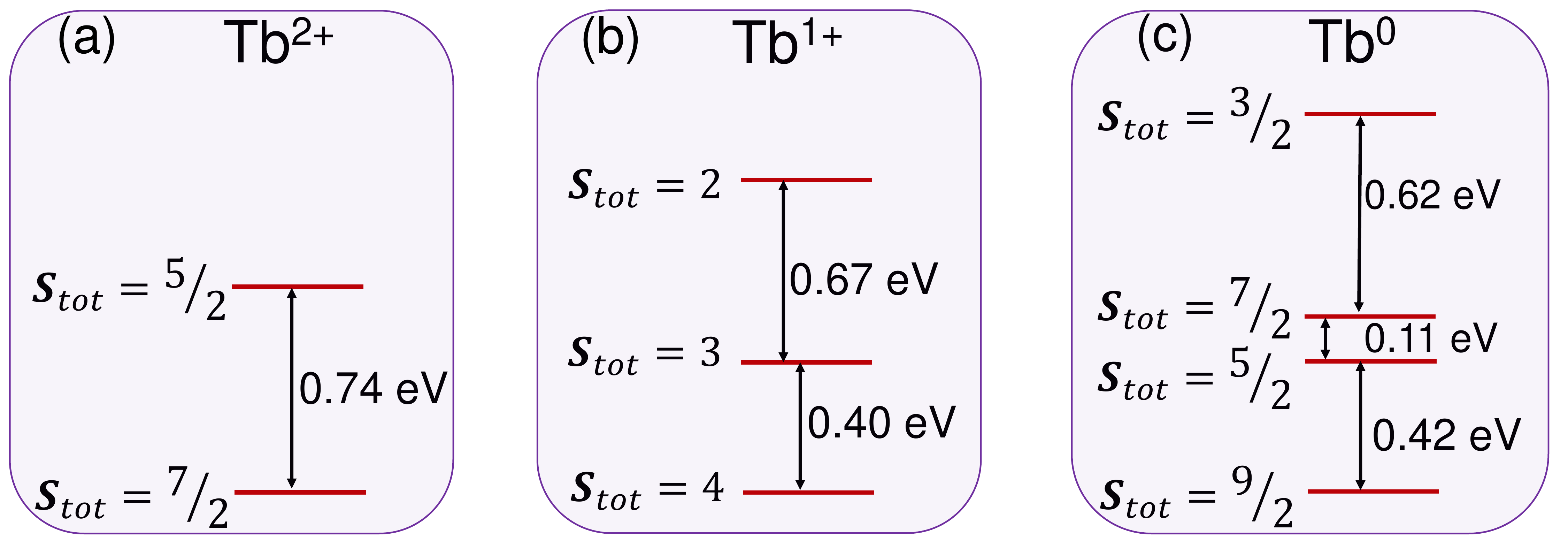}
\caption{Energy differences between the lowest spin-free levels for all considered spin states for (a) Tb$^{2+}$, (b) Tb$^{1+}$ and (c) Tb$^{0}$.} 
\label{Eex}
\end{figure}

The splitting of the ground multiplet by the CF and its response to the external magnetic field $\mathbf{B}$ can be described by the following effective spin Hamiltonian
\begin{equation}
\hat{H}=\sum_{k=2,4,6} \sum_{q=-k}^{k} B_k^q \hat{O}_k^q(\hat{\mathbf{S}})-\mu_B\mathbf{B}\cdot \mathbf{g}\cdot\hat{\mathbf{S}},
\label{CFHamiltonian}
\end{equation}
where $\hat{\mathbf{S}}$ is the pseudospin operator that for the Tb$^{3+}$ adatom corresponds to the total angular momentum operator for the ground multiplet ($J=6$). The first term is the CF Hamiltonian with $\hat{O}_k^q(\hat{\mathbf{S}})$ being $k$-th rank extended Stevens operators\cite{Rudowicz1985} and $B_k^q$ being corresponding CF parameters. Time reversal symmetry enforces only even integers $k$ and $q=-k,...,0,...,k$. The second term in Eq.~(\ref{CFHamiltonian}) is the Zeeman interaction that is described by the $g$-tensor, $\mathbf{g}$. The $\hat{O}_k^{q=0}$ operators commute with $\hat{S}_z$ and represent diagonal or uniaxial CF interactions, while $\hat{O}_k^{q \neq 0}$ operators do not commute with $\hat{S}_z$ and represent transverse or off-diagonal CF interactions. Due to the high symmetry of the adatom site ($C_{6v}$), the only nonzero transverse CF parameters is $B_6^6$. The nonzero CF parameters and the principal values of the $g$-
tensor are calculated for the ground multiplet (13 states) using the SINGLE ANISO formalism. The results are shown in Table~\ref{tab:2}. As seen, the $g$-
tensor is nearly isotropic and close to the ideal Lande $g$-factor value of 3/2. We also find that $B_6^6$ is much smaller than the diagonal CF parameters, and therefore, the CF is highly axial. This is a consequence of the high symmetry of the adatom site which requires all $\hat{O}_k^{q \neq 0}$ with $k<6$ being zero. As a result of the high axiality of the CF, $S_z$ (or the $z$ component of the total angular momentum for the ground multiplet, $J_z$) is approximately a good quantum number and the energy levels can be characterized by the magnitude of $J_z$. Indeed, as seen  in Fig.~\ref{Q3+Energy}, the spectrum is primarily formed by Ising doublets with a well defined $|J_z|$. The lowest energy doublet corresponds to $J_z=\pm6$ and it is separated from the first excited doublet ($J_z=\pm5$) by the zero-field splitting (ZFS) energy gap $E_{ZFS}=164$ cm$^{-1}$. For successive doublets, $|J_z|$ decreases by 1  until the highest lying singlet state with $J_z=0$. The Tb$^{3+}$ adatom has, thus, a magnetic moment $\mu=gJ\mu_B$ ($g\approx3/2$, $J=6$) and uniaxial anisotropy with easy axis perpendicular to the graphene layer. The two ground doublet states correspond to two opposite directions of the magnetic moment along the easy axis and are separated by a magnetic anisotropy barrier $E_{MAB}=569$ cm$^{-1}$.

The nonzero transverse CF parameter $B_6^6$ mixes different $J_z$ states as long as the difference between their $J_z$ values is an integer multiple of $q=6$. In particular, it couples $J_z=6$ and $J_z=-6$ states so that the ground doublet is split. This is known as tunnel splitting since it is responsible for the under-the-barrier quantum tunnelling of magnetization from $+J_z$ to $-J_z$ and vice versa. Since the $B_6^6$ parameter is very small and two multiples of $q=6$ (fourth order of perturbation theory) are required to couple $J_z=6$ and $J_z=-6$ states, the calculated tunnel splitting for the ground doublet is only $E_{TS}=10^{-4}$ cm$^{-1}$. Such low value indicates that quantum tunneling of magnetization for the ground doublet is very ineffective and is not responsible for magnetic relaxation of the Tb$^{3+}$ adatom. The $|J_z|=3$ is the only other doublet split by $B_6^6$. In this case $E_{TS}=2$ cm$^{-1}$ is significantly larger since the $J_z=3$ and $J_z=-3$ states are connected by a single multiple of $q=6$ (second order of perturbation theory). As a result, the thermal excitation (due to spin-phonon coupling) from the ground doublet to the third excited doublet on the same side of the barrier followed by the quantum tunneling of magnetization constitutes an efficient magnetic relaxation mechanism. This indicates that the effective magnetic anisotropy barrier corresponds to the energy of the third excited doublet, $U_{eff}=436$ cm$^{-1}$.

\subsection{Tb$^{2+}$ oxidation}

For the Tb adatom in the $2+$ oxidation state, an electron is added to the $Tb^{3+}$ $4f^8$ configuration. Calculations with large active space that includes Tb $4f$-, $6s$-, $5d$, and $6p$-like orbitals for different spin multiplicities revealed that for low-lying states the $4f^8$ configuration is always preserved and the extra electron occupies a $6s/5d_0$ orbital hybrid and the pair of $d_{xy}$, $d_{x2+y2}$  orbitals. Therefore, we proceed with an optimal active space CAS(9,10) that includes nine electrons in ten orbitals: seven $4f$, $6s/5d_0$, $d_{xy}$, and $d_{x2+y2}$. The active space is shown schematically in Fig.\ref{LevelDiagram}b. The active orbital pictures are also shown in Fig.~S2 in Supplementary Materials. Since there are again seven possible $4f^8$ configurations and the extra electron has three main configurations (due to three non-$4f$ orbitals), we used $3\times7=21$ roots in the SA-CASSCF calculations. 

Two different spin states are considered which mainly correspond to the extra electron having spin parallel or antiparallel to the $4f$ spin $S_{4f}=3$. For the parallel case the total adatom spin is $S_{tot}=S_{4f}+1/2=7/2$ (spin multiplicity is eight) while for the antiparallel case $S_{tot}=S_{4f}-1/2=5/2$ (spin multiplicity is six). The same active space (with the same type of active orbitals) and the number of roots are used for both spin states. We find that the $S_{tot}=7/2$ state has the lower energy and the $S_{tot}=5/2$ state is 0.74 eV higher (we compare the lowest roots for each spin state), see Fig.~\ref{Eex}a. This result indicates that there is a strong ferromagnetic intra-site exchange coupling between $4f$ spin $\hat{\mathbf{S}}_{4f}$ and the $6s5d$ spin $\hat{\mathbf{s}}$. Assuming that this interaction can be described by the Heisenberg-like Hamiltonian, $H_{ex}=-J_{ex}\hat{\mathbf{S}}_{4f}\cdot\hat{\mathbf{s}}$, the intra-site exchange parameter is $J_{ex}\approx0.2$ eV. 

\begin{figure}
\centering
\includegraphics[width=\linewidth]{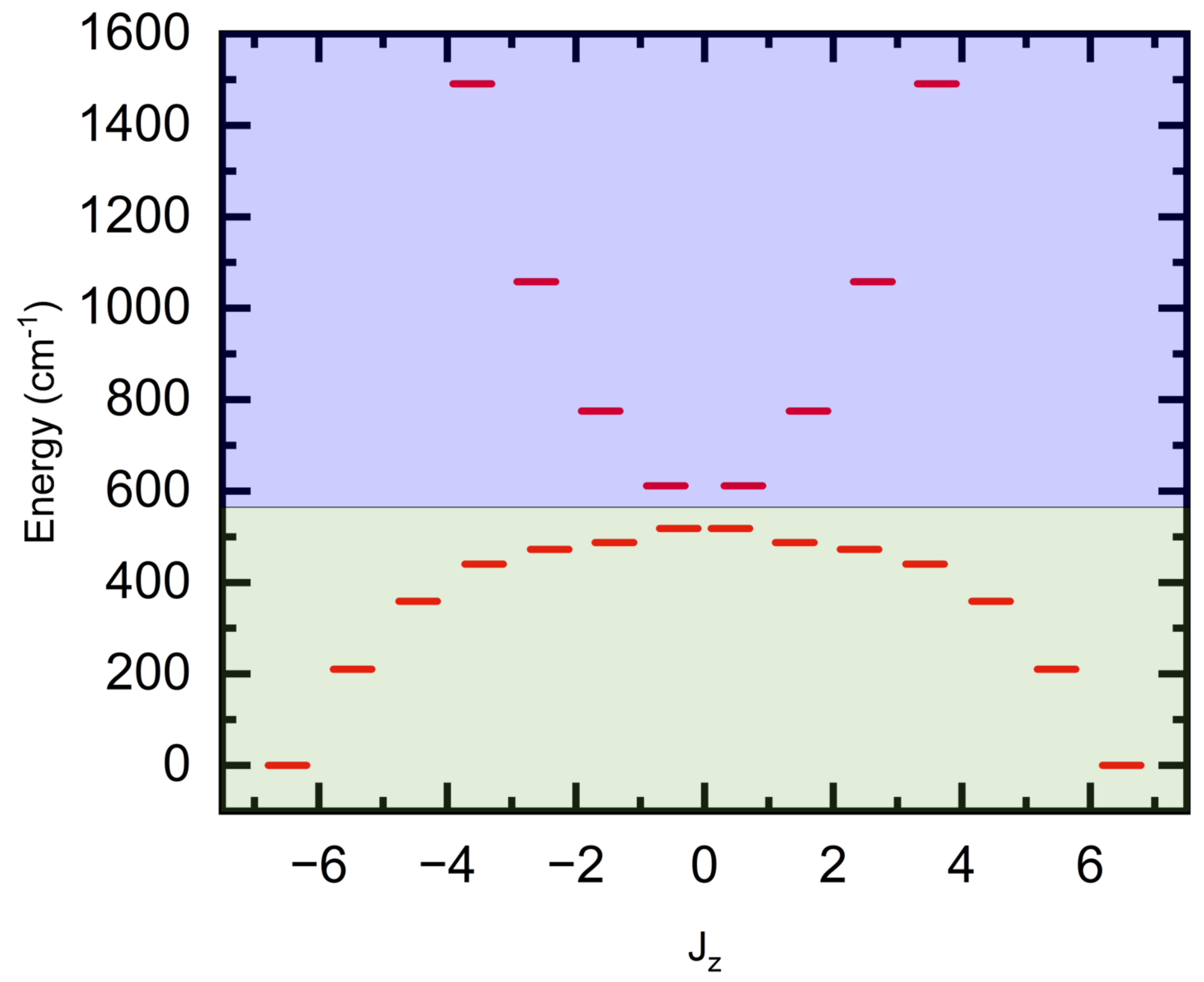}
\caption {Calculated low-energy magnetic states for the Tb$^{2+}$ adatom on graphene. The spectrum is  composed of Kramers doublets. Each doublet state is placed along the horizontal axis according to its value of the $z$ component of the total angular momentum.} 
\label{Q2+Energy}
\end{figure}

Energy levels of the Tb$^{2+}$ adatom are calculated by including SOC. The picture of $4f$ electron $J$-multiplets split by CF is, however, no longer valid due to strong interaction of the $4f$ electrons with the spin and orbital degrees of freedom of the non-$4f$ valence electron. This leads to a more complicated spectrum whose low-energy part is shown in Fig.~\ref{LevelDiagram}b (the numerical values are shown in Table S2 in Supplementary Materials). Here, we can identify two effective multiplets. The lower one (green background in Fig.~\ref{Q2+Energy}) consists of 14 states and corresponds to the effective spin $S=13/2$. The upper multiplet (purple background in Fig.~\ref{Q2+Energy}) is separated from the lower one by almost 100 cm$^{-1}$ and it corresponds to the effective spin $S=7/2$ (eight states). The upper multiplet is separated from higher lying states by about 775 cm$^{-1}$ (see Table S2 in Supplementary Materials).

CF parameters and the g-tensor are calculated for the lower multiplet with effective spin $S=13/2$ based on the Hamiltonian~(\ref{CFHamiltonian}) using the SINGLE ANISO formalism. The results are shown in Table~\ref{tab:2}. We find that the $g$-tensor is more anisotropic that in the Tb$^{3+}$ adatom case, but the principal values are still close to 3/2. Similarly as for the $3+$ oxidation state, CF is strongly uniaxial and $J_z$ is approximately a good quantum number. The $S=13/2$ multiplet is composed of Kramers doublet with perfect degeneracy as required by the time-reversal symmetry for systems with odd number of electrons (Kramers theorem). The Kramers doublets have a well defined magnitude of $J_z$ with larger $|J_z|$ values having the lower energy. In particular, the ground doublet corresponds to $J_z=\pm13/2$ and is separated from the first-excited doublet ($J_z=\pm11/2$) by $E_{ZFS}=210$ cm$^{-1}$. These results indicate that at low energies the Tb$^{2+}$ has a magnetic moment $\mu=gS\mu_B$ ($g\approx3/2$, $J=13/2$) and uniaxial anisotropy with easy axis perpendicular to the graphene layer. The height of the magnetic anisotropy barrier is $E_{MAB}=518$ cm$^{-1}$.

The small transverse CF interaction ($B_6^6$) again mixes $J_z$ states for which the difference in $J_z$ values is an integer multiple of $q=6$. These interactions, however, do not lead to doublet splittings since the doublet degeneracy is ensured by the time reversal symmetry. Therefore, for Kramers systems, the transverse CF interactions by itself do not lead to quantum tunneling of magnetization. For most of real nanomagnets including magnetic adatoms, however, there are always small transverse magnetic fields due to various magnetic impurities or nuclear spins. Such transverse fields can cause quantum tunneling between two partner states of a Kramers doublet provided that the transverse component of the effective magnetic moment of the doublet is nonzero\cite{SMMbook}. Principal values of the $g$-tensor for all Kramers doublets of the $S=13/2$ multiplet are shown in Table~S5 in Supplementary Materials. As seen, for the three lowest doublets the transverse components of the $g$-tensor (and, thus, the effective magnetic moment) are very small so the quantum tunneling is very ineffective for these doublets. On the other hand, for the third excited doublet the transverse $g$-factors are significant ($\approx0.7$) and efficient quantum tunneling is expected. Therefore, we identify the temperature assisted quantum tunneling of magnetization through the third-excited doublet as an efficient magnetic relaxation mechanism. The energy of the third-excited doublet determines, thus, the effective magnetic anisotropy barrier $U_{eff}=440$ cm$^{-1}$.

\subsection{Tb$^{1+}$ oxidation}

\begin{figure}
\centering
\includegraphics[width=\linewidth]{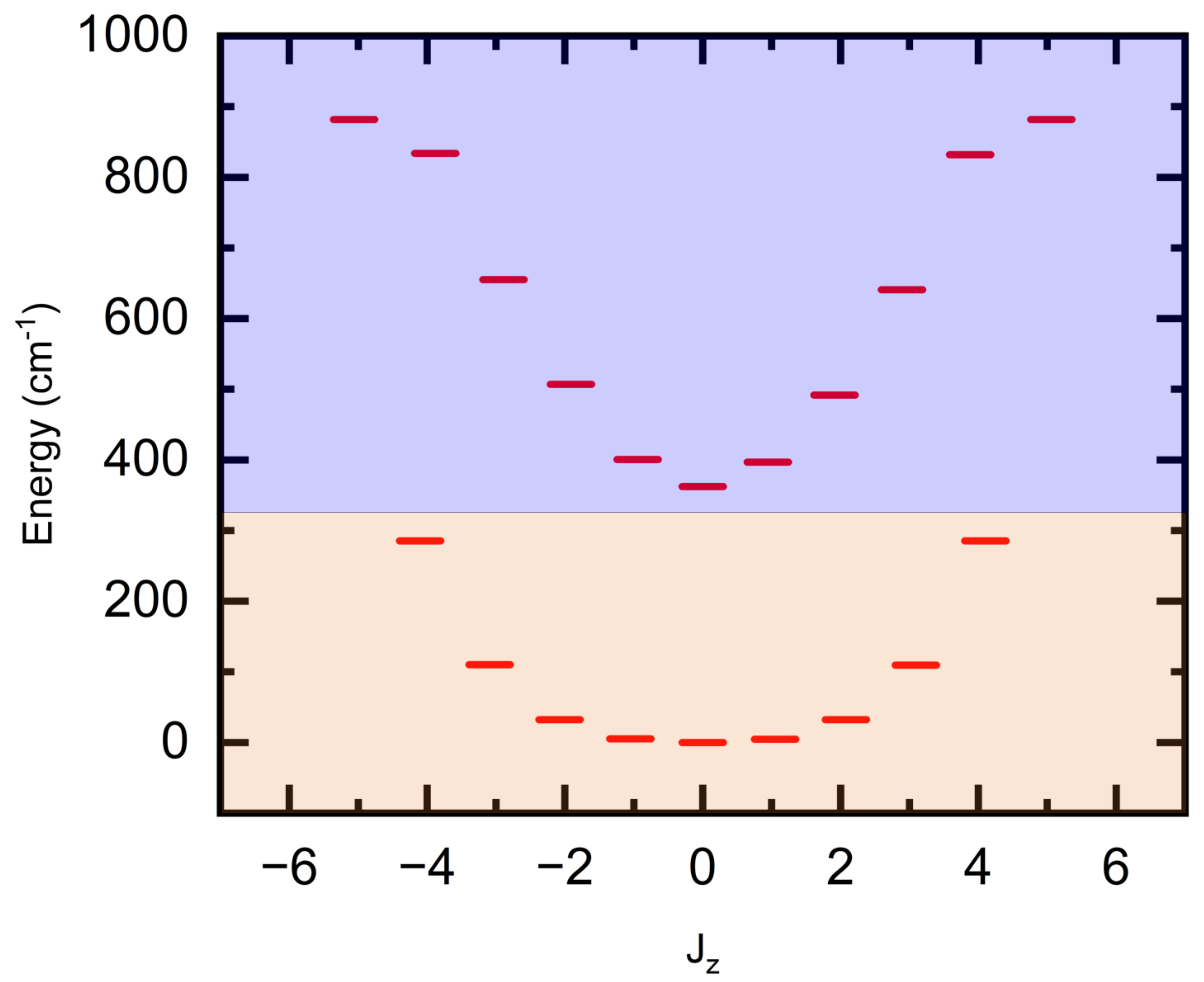}
\caption {Calculated low-energy magnetic states for the Tb$^{1+}$ adatom on graphene. The spectrum is primarily  composed of (quasi-)doublets. Each doublet state is placed along the horizontal axis according to its value of the $z$ component of the total angular momentum. The singlet states have $J_z=0$.} 
\label{Q1+Energy}
\end{figure}

For the Tb adatom in the $1+$ oxidation state, two additional electrons are added to the $Tb^{3+}$ $4f^8$ configuration. Following the same approach as for $Tb^{2+}$ we find that for low-lying states the $4f^8$ configuration is essentially preserved, one of the extra electrons occupies a $6s/5d_0$ orbital hybrid, and the other extra electrons occupies $d_{xy}$ and $d_{x2+y2}$  orbitals, see Fig.~\ref{LevelDiagram}c.  We, therefore, choose the CAS(10,10) active space with ten electrons and the same type of active orbitals as for the Tb$^{2+}$ case. The active orbital pictures are shown in Fig.~S3 in Supplementary Materials. As there are two main configurations of the non-$4f$ electrons and seven possible $4f^8$ configurations, the number of roots used in the SA-CASSCF calculations is $2\times7=14$. 

We consider three different spin states $S_{tot}=4$, $S_{tot}=3$, and $S_{tot}=2$. The first one corresponds primarily to both non-$4f$ electrons having spins parallel to the $4f$ electron spin of $S_{4f}=3$. For the two other spin states we also mostly have $S_{4f}=3$ with one or two of non-$4f$ electrons having spins antiparallel to the $4f$ spin. We find that $S_{tot}=4$ has the lowest energy with the $S_{tot}=3$ state being 0.4 eV higher in energy and the $S_{tot}=2$ state lying additional 0.67 eV higher (see Fig.~\ref{Eex}b). These results are again indicative of a strong intra-site exchange coupling between $4f$ and $6s5d$ spins. However, pair exchange parameters between different types of orbitals cannot be unambiguously extracted from our results.

Low-energy electronic spectrum calculated by including SOC is shown in Fig.~\ref{Q1+Energy} (numerical values are shown in Table S3 in Supplementary Materials). We can again identify two multiplets, but similarly as for the $2+$ oxidation states, they do not correspond to the $J$-multiplets typical for trivalent lanthanide ions. Instead, these are effective multiplets that originate from the interaction between $4f$ states with the spin and orbital degrees of freedom of $6s5d$ electrons in the presence of SOC. The lower multiplet (orange background in Fig.~\ref{Q1+Energy}) is composed of nine states and it corresponds to the effective spin of $S=4$. The upper multiplet (purple background in Fig.~\ref{Q1+Energy}) is only 77 cm$^{-1}$ above the lower one and it corresponds to the effective spin $S=5$ (11 states). The higher lying states are about 200 cm$^{-1}$. It is important to point out that the identification of effective multiplets and their effective spins is somewhat ambiguous here due to relatively small gaps between the multiplets.

Using the SINGLE ANISO formalism we evaluated $g$-tensor and CF parameters for the lower effective multiplet based on the Hamiltonian~(\ref{CFHamiltonian}), see Table~\ref{tab:2}. We find that the transverse $g$-factors ($g_1$ and $g_2$) are close to 2.5 and the $g$-factor along the direction perpendicular to the graphene plane is about 2. This significant anisotropy of the $g$-factors is caused by interaction with the upper multiplets that lie relatively close in energy. The CF is again strongly axial and $J_z$ is approximately a good quantum number with the energy levels depending on $|J_z|$. The spectrum is, thus, composed of doublets except for the $J_z=0$ state that is a singlet. Since for Tb$^{1+}$ we have an integer number of electrons, these are Ising doublet whose degeneracy is not protected by the time reversal symmetry. In fact, the transverse CF interaction ($B^6_6$) and the interaction with the higher multiplets mix different $J_z$ states and leads to significant splitting of the doublets (see Table S3 in Supplementary Materials). Most importantly, we find that the dominant uniaxial parameter, $B^2_0$ is positive. This indicates that the Tb$^{1+}$ adatom has an in-plane magnetic anisotropy. Indeed, the ground energy level is the $J_z=0$ singlet and the first-excited doublet is only about 5 cm$^{-1}$ above. Stable magnetic moment is, thus, not expected for the Tb$^{1+}$ adatom.

\subsection{Tb$^{0}$ oxidation}

\begin{figure}[b]
\centering
\includegraphics[width=\linewidth]{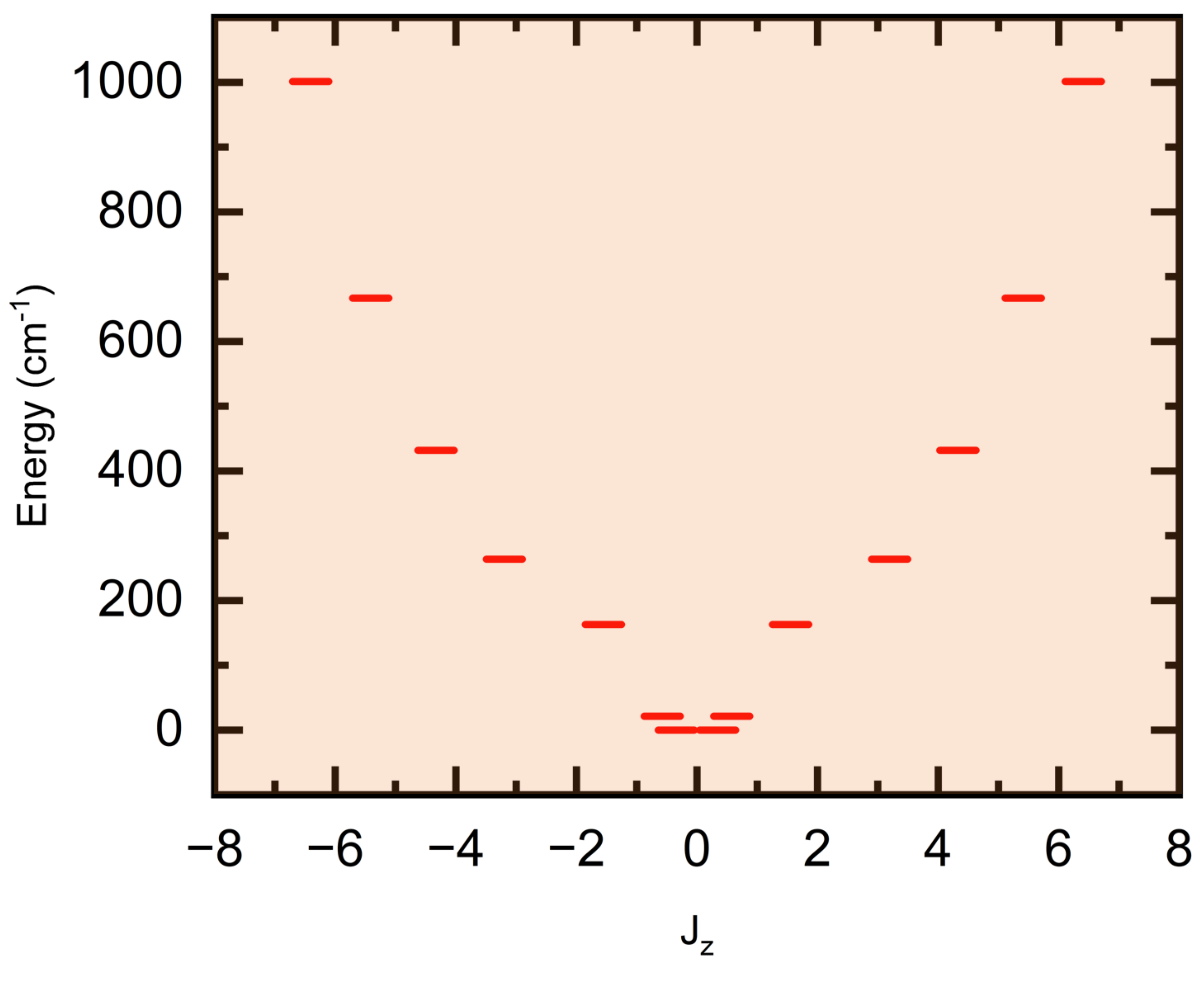}
\caption{Calculated low-energy magnetic states for the Tb$^{0}$ adatom on graphene. The spectrum is  composed of Kramers doublets. Each doublet state is placed along the horizontal axis according to its value of the $z$ component of the total angular momentum.} 
\label{Q0Energy}
\end{figure}

For the neutral Tb adatom our calculations show that the $4f^8$ configuration is still retained and the three additional valence electrons occupy $6s/5d_0$, $d_{xy}$, and $d_{x2+y2}$ orbitals, see Fig.~\ref{LevelDiagram}d. The optimal active is, thus, CAS(11,10) with eleven active electrons and the same type of active orbitals as for the other non-$3+$ oxidation states. The active orbital pictures are shown in Fig.~S4 in Supplementary Materials. Since there is only a single configuration of the non-$4f$ electrons, we use seven roots in the SA-CASSCF calculations to account for seven possible $4f^8$ configurations.

Four different spin states are considered including $S_{tot}=9/2$, $S_{tot}=7/2$, $S_{tot}=5/2$, and $S_{tot}=3/2$. In general, for all spin state we mostly have $S_{4f}=3$ and different spin states differ by the number of the non-$4f$ electrons with spins parallel to the $4f$ spin. As expected, due to strong intra-site exhcange coupling, $S_{tot}=9/2$ and $S_{tot}=3/2$ states have the lowest and the highest energy, respectively (see Fig.~\ref{Eex}c). The energy difference between these two states is around 1.1 eV. The other two states lie in between these two, but interestingly the $S_{tot}=5/2$ state has a lower energy than that of $S_{tot}=7/2$ state.

The low-lying energy levels for the Tb$^{0}$ adatom on graphene calculated by including SOC are shown in Fig.~\ref{Q0Energy}. Due to absence of orbital degrees of freedom for the $6s5d$ electrons and strong intra-site exchange coupling, the electronic spectrum is expected to originate from the total electronic spin $S_{tot}=9/2$ and the $4f$ electrons orbital angular momentum $L_{4f}=3$ which interact with each other by SOC. This leads to a multiplet structure where the multiplets are characterized by the total angular momentum of the $4f$ and $6s5d$ electrons, $J_{tot}$. Such physical picture has been used to explain electronic structure of SMMs based on divalent lanthanides where a single non-$4f$ electron occupies a nondegenerate $6s/5d_0$ orbital\cite{Dubrovin2019,Jin2023}. It was found that in these systems the Hund's 3rd rule was satisfied and the ground multiplet corresponded to the maximum possible value of $J_{tot}$\cite{Zhang2020}. In our case, however, the ground multiplet corresponds to $J_{tot}=13/2$ while the maximum value of $J_{tot}$ is $L_{4f}+S_{tot}=15/2$. This indicates that the Hund's 3rd rule cannot always be extended to lanthanides atoms with both $4f$ and $6s5d$ electrons.

The $J_{tot}$-multiplets are split by CF into Kramers doublets (as expected for the Tb$^{0}$ adatom with an odd number of valence electrons). This splitting and the interaction with the external magnetic field can be described by the effective spin Hamiltonian (\ref{CFHamiltonian}). We calculate the $g$-factors and CF parameters for the $J_{tot}=13/2$ multiplet (see Table~\ref{tab:2}). Even though the multiplet is separated from higher levels by almost 700 cm$^{-1}$, the $g$-factors show some anisotropy with the transverse and $z$ components of the $g$-tensor being about 1.9 and 1.7, respectively. The CF is strongly axial due to high symmetry of the adatom site. Therefore, $J_z$ is an approximately good quantum number and the Kramers doublets are characterized by $|J_z|$. We find that energy increases with $|J_z|$. Thus, we have in-plane magnetic anisotropy that is caused by the dominant $B^0_2$ CF parameter being positive. While this result may suggest that the Tb$^{0}$ adatom on graphene has no stable magnetic moment, we point out that the magnetic moment in the ground doublet ($J_z=\pm1/2$) is nonzero and the slow magnetic relaxation has been observed for Dy adatom on SrTiO$_3$ with an in-plane anisotropy\cite{Bellini2022}.

\section{Conclusions}

We investigate low-energy spectrum and magnetic interactions for Tb adatom on graphene from first principles using combination of DFT and CASSCF-RASSI-SOC quantum chemistry calculations. Different adatom oxidation states are considered including Tb$^{3+}$, Tb$^{2+}$, Tb$^{1+}$, and Tb$^{0}$. The key findings are as follows:

\begin{itemize}

\item The graphene hollow site with the $C_{6v}$ symmetry is found to be the preferred adsorption site. 

\item For all oxidation states, the Tb $4f^8$ configuration is found with other adatom valence electrons occupying $5d_{xy}$, $5d_{x2+y2}$, and $6s/5d_0$ single-electron orbitals.

\item The spins of the $5d/6s$ electrons are parallel to the $4f$ spin due to large intra-site exchange coupling that is significantly stronger than SOC.

\item For Tb$^{3+}$ the low-energy spectrum arises from the $J=6$ multiplet being split by the graphene CF but still well separated in energy from other $J$-multiplets. For other oxidation states the interaction of $4f$ electrons with spin and orbital degrees of freedom of $6s5d$ electrons in the presence of SOC leads to formation of effective multiplets that are split by the graphene CF. The effective multiplets lie relatively close in energy and can interact with each other.

\item For Tb$^{3+}$ and Tb$^{2+}$, we find uniaxial magnetic anisotropy and identify the temperature assisted quantum tunneling of magnetization through the third-excited doublet as the dominant magnetic relaxation mechanism. The corresponding effective anisotropy barriers are about 440 cm$^{-1}$.

\item For $Tb^{1+}$ and $Tb^{0}$, we find in-plane magnetic anisotropy. This result indicates that occupation of rare earth adatom $5d6s$ orbitals can dramatically change the CF that acts on the $4f$ electrons. 

\end{itemize}

\section*{Acknowledgements}
This work is supported by the NSF EPSCoR Cooperative Agreement OIA-2044049, Nebraska's EQUATE collaboration. We acknowledge the University of Nebraska Holland Computing Center for computational resources.

%

\end{document}